\documentstyle[emulateapj,psfig]{article}

\submitted{Received 2001 March 16; accepted 2001 April 5}

\lefthead{Halpern {\it et al.}}
\righthead{\psr : An Energetic Young Pulsar}

\begin{document}

%
%

\def\source{3EG~J2227+6122}
\def\ro{{\it ROSAT\/}}
\def\asca{{\it ASCA\/}}
\def\xray{RX/AX~J2229.0+6114}
\def\vla{VLA~J2229.0+6114}
\def\psr{PSR~J2229+6114}

\title{\psr : Discovery of An Energetic Young Pulsar in the
Error Box of the EGRET Source 3EG J2227+6122}

\author{J. P. Halpern, F. Camilo, E. V. Gotthelf,
and D. J. Helfand}
\affil{Columbia Astrophysics Laboratory, Columbia University,
550 West 120th Street, New York, NY 10027}
\authoremail{jules@astro.columbia.edu}
\author{M. Kramer, A. G. Lyne}
\affil{University of Manchester, Jodrell Bank Observatory, Macclesfield,
Cheshire, SK11 9DL, UK}
\author{K. M. Leighly}
\affil{Department of Physics and Astronomy, The University of Oklahoma,
440 W. Brooks St., Norman, OK 73019}
\and
\author{M. Eracleous}
\affil{Department of Astronomy and Astrophysics, The Pennsylvania State
University, \\525 Davey Laboratory, University Park, PA 16802}

\begin{abstract}
\rightskip 0pt \pretolerance=100 \noindent
We report the detection of radio and X-ray pulsations at a period of
51.6 ms from the X-ray source \xray\ in the error box of the EGRET
source \source.  An ephemeris derived from a single \asca\ observation
and multiple epochs at 1412~MHz from Jodrell Bank indicates steady
spin-down with $\dot P = 7.83 \times 10^{-14}$ s~s$^{-1}$.  From the
measured $P$ and $\dot P$ we derive spin-down power $\dot E = 2.2
\times 10^{37}$ ergs~s$^{-1}$, magnetic field $B_{\rm p} = 2.0 \times
10^{12}$~G, and characteristic age $P/2\dot P = 10,460$~yr.  An image
from the {\it Chandra} X-ray Observatory reveals a point source
surrounded by centrally peaked diffuse emission that is contained
within an incomplete radio shell.  We assign the name G106.6+2.9 to
this new supernova remnant, which is evidently a pulsar wind nebula.
For a distance of 3~kpc estimated from X-ray absorption,
the ratio of X-ray luminosity to spin-down power is
$\approx 8 \times 10^{-5}$, smaller than that of most pulsars, but
similar to the Vela pulsar.  If \psr\ is the counterpart of
\source\ then its efficiency of gamma-ray production, if isotropic, is
$0.016\,(d/3\,{\rm kpc})^2$. It obeys an established trend of $\gamma$-ray
efficiency among known $\gamma$-ray pulsars which, in combination with
the demonstrated absence of any other plausible counterpart for
\source, makes the identification compelling.  If confirmed, this
identification bolsters the pulsar model for unidentified Galactic
EGRET sources.
\end{abstract}

\section{Introduction}

\source\ (Hartman et al. 1999)
is one of a number of ``unidentified'' EGRET sources at low
Galactic latitude, $(\ell,b) = (106.\!^{\circ}6,2.\!^{\circ}9)$, for
which a pulsar origin is the hypothesis favored by many
authors (e.g., Halpern
\& Ruderman 1993; Kaaret \& Cottam 1996; Yadigaroglu \& Romani 1997).
Emission attributed to this source was also detected by COMPTEL
(Iyudin et al. 1997), and probably by {\it COS B} (Wills et al. 1980).
We recently presented an X-ray, radio, and optical study of the error
circle of \source\ that revealed only one plausible candidate, a highly
polarized, flat-spectrum radio shell superposed on a
compact, non-thermal X-ray source with power-law photon index $\Gamma =
1.51 \pm 0.14$ and with no obvious optical counterpart (Halpern et al.
2001, hereafter Paper~I).  We concluded that the most likely
interpretation was a young pulsar and associated wind nebula (PWN).  The
distance was estimated as 3~kpc from the X-ray absorption column density
of $(6.3 \pm 1.3) \times 10^{21}$~cm$^{-2}$.  At that distance, the spin-down
luminosity required to power the X-ray and $\gamma$-ray luminosities of
$1.7 \times 10^{33}$ and $3.7 \times 10^{35}$ ergs~s$^{-1}$,
respectively, would be similar to that of the younger known
EGRET pulsars.  In this Letter, we report follow-up observations that
confirm a pulsar origin for \xray.

\section{Chandra Image}

We observed the X-ray source \xray\ for 17,728~s on 2001 February 14
with the {\it Chandra} imaging CCD array ACIS--I.  The target was offset
by $0.\!^{\prime}5$ from the default ACIS--I pointing position in each
direction in order to avoid the inter-CCD gaps.  A point source was
clearly detected, surrounded by diffuse emission
with a centrally peaked morphology (Figure~1).
There is evidently an arc of emission that resembles
similar structures in the Vela nebula (Helfand, Gotthelf, \& Halpern
2001), surrounded by fainter diffuse
emission that is largely confined within the boundaries of the radio
shell, which has a radius of $\approx 100^{\prime\prime}$.
The radial profile of the X-ray flux is shown
in Figure~2. The central point
source accounts for $29\%$ of the total flux within
$100^{\prime\prime}$. 
The total 2--10~keV flux is $1.3 \times
10^{-12}$~ergs~cm$^{-2}$~s$^{-1}$, only slightly smaller than the value of
$1.56 \times 10^{-12}$~ergs~cm$^{-2}$~s$^{-1}$ that we found in Paper~I
using \asca.  It is possible that the \asca\ GIS, with its large
point-spread function, encompassed some extended diffuse emission that is
not easily recovered in this short {\it Chandra} observation.
In combination with the non-thermal spectrum of the X-ray nebula,
the morphology indicates a ``composite'' supernova remnant
(to which

\bigskip
\centerline{
\psfig{file=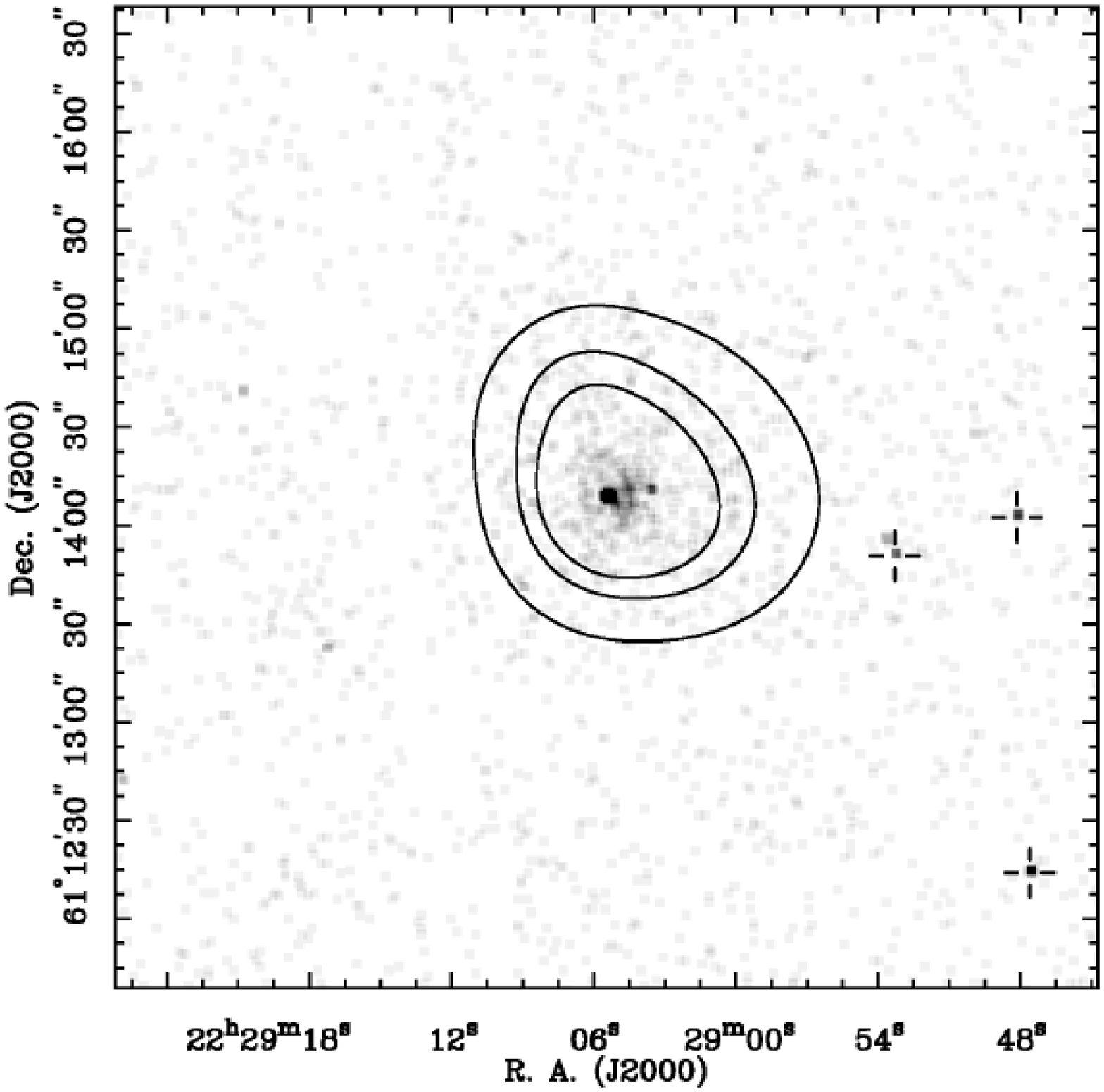,angle=270,height=3.2in,clip=}
}
\medskip
{\footnotesize FIG. 1. --- A portion of the {\it Chandra} ACIS--I image
showing \xray.  The
image is binned into $\approx 1^{\prime\prime}$ pixels and smoothed
with a 3-pixel top-hat filter.  It is displayed with square-root
intensity scaling with the contrast set to highlight faint diffuse
structures.  The linearly-spaced overlaid contours correspond to an
adaptively smoothed map. Also shown are the locations of X-ray point
sources detected with $> 2\sigma$ confidence.}
\bigskip

\noindent
we assign the name G106.6+2.9), although the radio shell
is probably a shock between
the PWN and the surrounding medium rather than a
supernova blast wave (see \S 6).
No thermal emission has yet been
detected from this remnant at X-ray or optical wavelengths.

\bigskip
\centerline{
\psfig{file=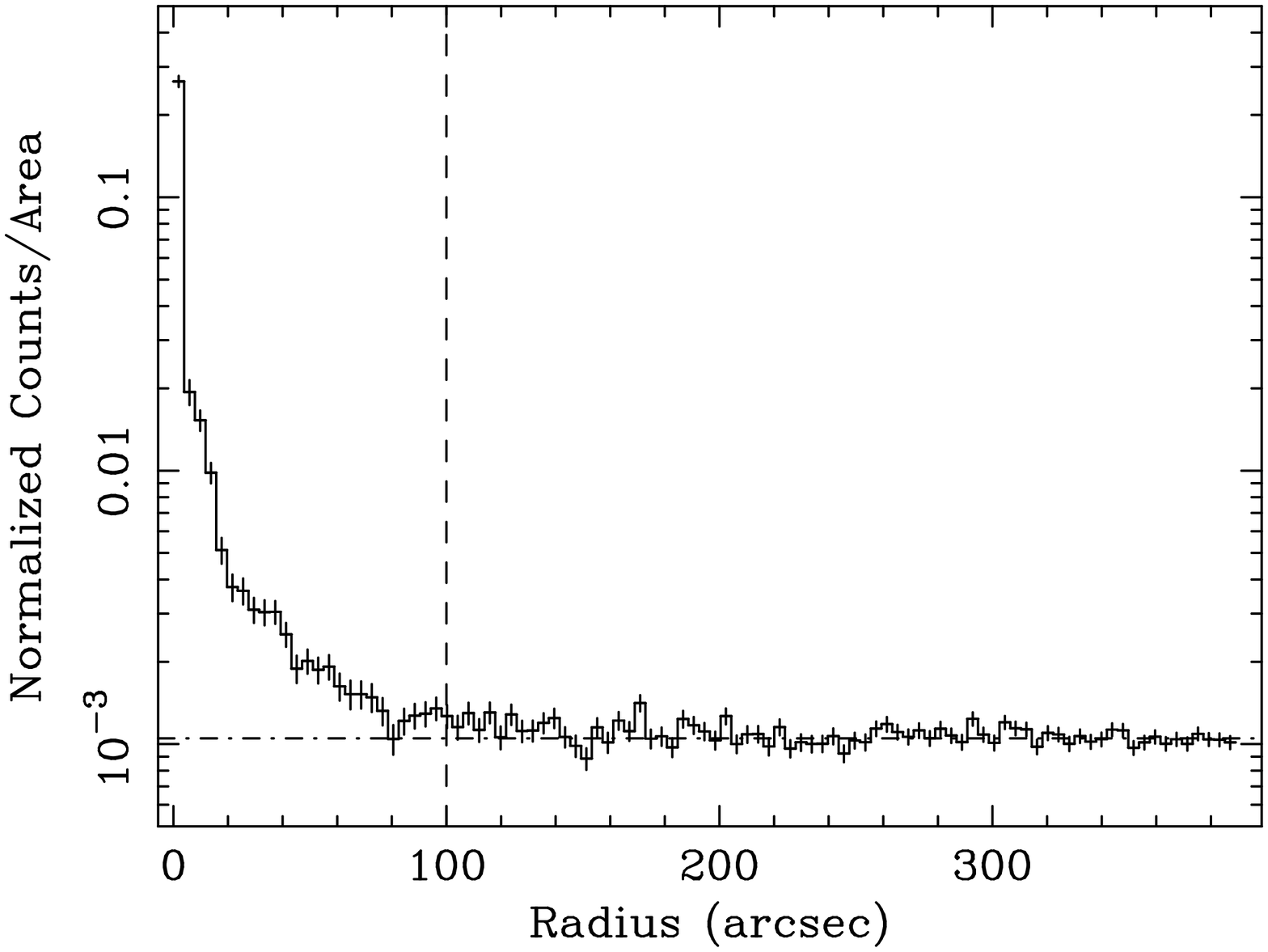,angle=270,height=2.5in,clip=}}
\medskip
{\footnotesize FIG. 2. --- Radial profile of the {\it Chandra} ACIS--I
image of \xray.  Each radial bin is 8 ACIS pixels wide
($3.\!^{\prime\prime}9$). The profile is normalized by area and
corrected for the instrument response across the focal plane, including
mirror vignetting and inter-CCD chip gaps. The pulsar flux is confined
to the central bin, around which is clearly evident at least two
components of enhanced emission, one inside and one outside of
$20^{\prime\prime}$.  The one-sigma error bar is shown for each data
point.  The {\it dot-dashed line} represents the estimated background level.
The {\it dashed line} denotes the approximate outer boundary of the
radio shell from Paper~I.}
\bigskip

There are eight X-ray sources in the ACIS image that can be immediately
identified by inspection with bright
stars on the Palomar Observatory
Sky Survey.  Two of these are present on our optical CCD images
of the field that were discussed in Paper~I.  We used them to derive a
correction of $1.\!^{\prime\prime}5$ to the X-ray aspect solution,
obtaining the precise position for the point source
listed in Table~1,
with an estimated uncertainty radius of $0.\!^{\prime\prime}5$.

\section{Radio Pulsar}

The position of \xray\ was observed with the 76m Lovell radio telescope
at Jodrell Bank on 2001 February 27 and 28 for 1.5 and 2.4\,hr,
respectively.  The observations used a cryogenic receiver at a central
frequency of 1412\,MHz and a filter bank spectrometer with 64 channels,
each 1\,MHz wide, for each of two orthogonal polarizations.  After
square-law detection and summing of complementary polarizations, the
total-power levels in each of 64 channels were integrated every 1\,ms,
1-bit digitized, and written to tape for off-line analysis.

Standard search algorithms detected an unmistakable pulsed signal at a
barycenter-corrected period $P=51.6235$~ms and dispersion measure DM
$=200\pm10$ cm$^{-3}$~pc with signal-to-noise ratio (S/N) of 13.9 and
18.6 on February 27 and 28, respectively.  These S/N values are
consistent with each other after scaling for the relative integration
times, suggesting that the pulsar does not scintillate markedly.  Based
on this and the known telescope sensitivity, we estimate that the
period-averaged flux density of the pulsar at this frequency is
0.25\,mJy, with $\sim 30\%$ uncertainty.  This is a rather low value,
fully consistent with its non-detection in previous ``all-sky''
surveys, and with the median rms of 0.12\,mJy reported for the VLA
image in Paper~I.  The radio pulse profile (Figure 3) consists of one
featureless peak with full-width at half-maximum of $0.08\pm0.02$
cycles.

We began regular timing observations of this source, and a
phase-connected solution to times-of-arrival spanning 11
days in 2001 March yields the period listed in Table~1, and
$\dot P = (7.80 \pm 0.03) \times 10^{-14}$ s~s$^{-1}$.
The more accurate $\dot P$

\bigskip
\centerline{
\psfig{file=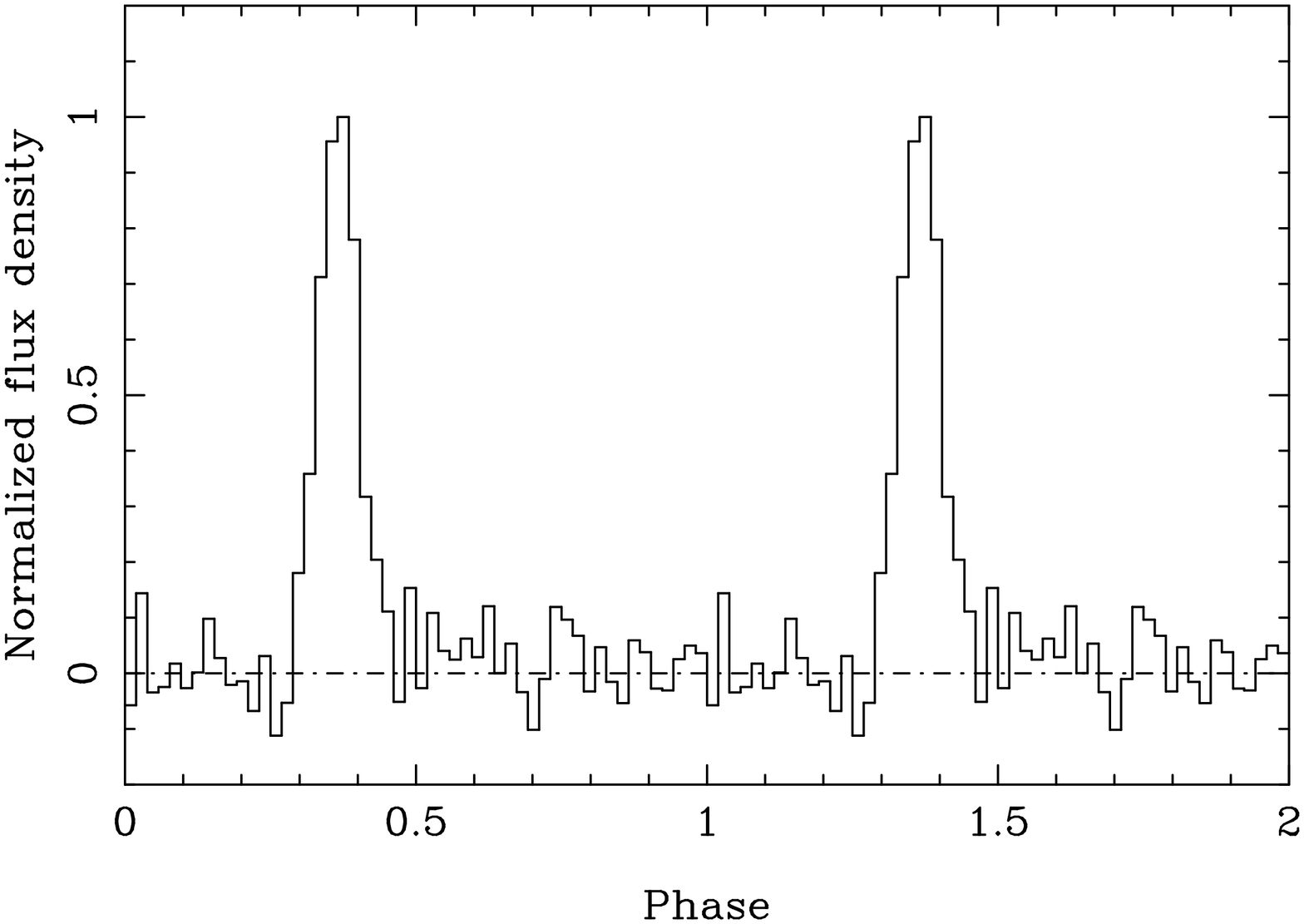,angle=270,height=2.4in,clip=}}
\medskip
{\footnotesize FIG. 3. --- Radio pulse profile of \psr\ at 1412~MHz. 
The instrumental resolution is $\approx 0.02$ of the period.
Phase zero is arbitrary.}
\bigskip

\noindent
listed in Table~1 is derived from a fit to this
period in 2001 March and the
period subsequently found in the
X-ray data obtained 1.6 yr earlier (\S 4).
Thus far, our data 
do not meaningfully constrain the amount of timing noise or glitch
activity that may be present in \psr.
If we assume a distance of 3~kpc as estimated from the X-ray spectrum
in Paper~I, the free electron/distance model of Taylor \& Cordes~(1993)
would predict DM $=75$ cm$^{-3}$~pc, significantly less than the
observed DM $=200$. Conversely, for the observed DM the model predicts
a distance of 12~kpc.  It is known that for individual objects the
model distances/DMs can be wrong by a factor of a few, and this may be
the case for \psr.  In fact, the rule of thumb of 0.1 free
electrons per hydrogen atom, combined with the measured DM, yields a
predicted $N_{\rm H} = 6\times10^{21}$\,cm$^{-2}$, in exact agreement
with the X-ray measured value from the \asca\ spectrum.  While the
conversion from column density to distance is itself uncertain, we
tentatively retain the previous ``X-ray'' estimate of $d=3$~kpc,
while acknowledging that considerable uncertainty remains.

\section{X-ray Pulsar}

Following the radio pulsar discovery in 2001 February, we searched the
113,700 s of \asca\ GIS data obtained on 1999 August 4--7 for
pulsations at periods slightly shorter than the radio, incorporating in
the folding the $\dot P$ that would be necessary to connect to the
radio period obtained 1.6~yr later.  Using the $Z^2_n$ test (Buccheri
et al. 1983) to gain sensitivity to $n$ harmonics of the fundamental
period, we found an unambiguous signal with $Z^2_7 = 99.4$ at
$P = 51.6196134 \pm 0.0000005$~ms
(MJD 51394.365), requiring $\dot P = 7.827 \times
10^{-14}$~s~s$^{-1}$ to connect to the radio ephemeris.  The $Z^2_n$
statistic is distributed as $\chi^2$ with $2n$ degrees of freedom.  The
probability of a spurious detection is $6 \times 10^{-15}$ per trial,
while only $\sim 10^4$ independent trial periods were searched.
Figure 4 shows
the folded light curve in the 0.8--10 keV band, extracted from a region of
radius $4^\prime$ around the source.  There are evidently two pulses
approximately $180^{\circ}$ apart in phase, although they are not
symmetric.  The rise time of the higher pulse is unresolved; the bin
size in Figure~4 (2.55~ms) is comparable to the instrumental time
resolution.  (Approximately half the data were obtained with 0.5~ms
resolution, and half with 3.9~ms resolution.)  Since most of the
extracted flux is diffuse, it is not possible to measure the true
pulsed fraction accurately.  After subtracting background from
an annulus surrounding the nebula, the \asca\ GIS light curve has
a pulsed fraction of 22\%, but
since the {\it Chandra} image indicates that only 29\% of these photons
are from the pulsar, the true pulsed fraction must be at least 75\%.

\section{Optical Observations}

Optical images of the location of \psr\ from the MDM 2.4m telescope
were shown in Figure~3 of
Paper~I.  The precise {\it Chandra} position of \psr\
is now seen to fall only
$1.\!^{\prime\prime}5$ north of the $R = 17.3$ star marked A in that
figure, and a similar distance south of a fainter star of $R = 21.3$.
Given the uncertainty of only $0.\!^{\prime\prime}5$, the X-ray
position is inconsistent with either of these stars.  Star A is, in
fact, a highly reddened A star as we determined using the KPNO 2.1m
telescope and Goldam

\bigskip
\centerline{
\psfig{file=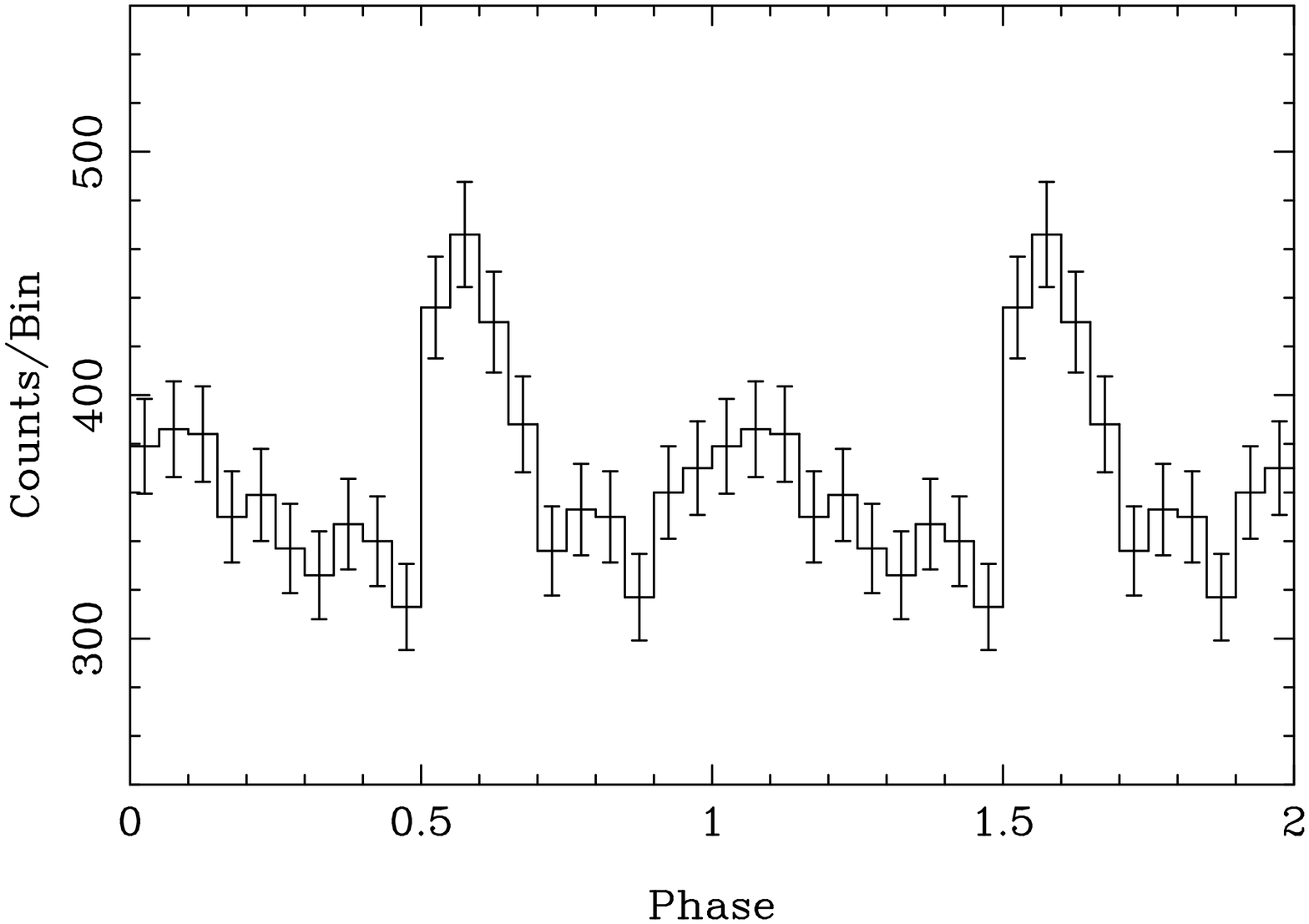,angle=270,height=2.5in,clip=}}
\medskip
{\footnotesize FIG. 4 --- X-ray pulse profile of \psr\ in the 0.8--10 keV
band from the \asca\ GIS.  The instrumental time resolution is
comparable to the width of one phase bin.  Phase zero is arbitrary.}
\bigskip

\noindent
spectrograph.  The proximity to this bright star
will make any further optical search for the pulsar difficult from the
ground.  Although our images reach a limiting magnitude of $R = 24.5$
in good seeing, a more conservative limit of $R > 23$ probably applies
at the location of the pulsar in the wings of the bright star.

\section {Discussion}

Applying the standard magnetic dipole model for rotation powered
pulsars, we can use the measured $P$ and $\dot P$ of \psr\ to derive
the spin-down power $\dot E = I\Omega\dot\Omega = 2.2 \times 10^{37}$ 
ergs~s$^{-1}$, the
magnetic field $B_{\rm p} = 3.2 \times 10^{19}\,(P\dot P)^{1/2} = 
2.0 \times 10^{12}$~G, and the
characteristic age $\tau = P/2\dot P = 10,460$~yr.
Compared to known $\gamma$-ray pulsars \psr\ is
second only to the Crab in spin-down power, and it is significantly more
luminous than the Vela pulsar (PSR B0833--45), which has $\dot E = 6.9
\times 10^{36}$ erg~s$^{-1}$.
For a distance of 3~kpc estimated from X-ray
absorption and radio pulse dispersion, the ratio of total 2--10 keV
X-ray luminosity to spin-down power is $\approx 8 \times 10^{-5}$, smaller
than that of most pulsars, but similar to Vela (Helfand et al. 2001).
If \psr\ is the counterpart of \source, then
its luminosity above 100~MeV is $\approx 3.7 \times 10^{35}$
ergs~s$^{-1}$, and its efficiency of gamma-ray production, if
isotropic, is $0.016\,(d/3\,{\rm kpc})^2$.
For Vela, the same analysis gives a $\gamma$-ray
efficiency of $0.03\,(d/500\,{\rm pc})^2$.

Among the pulsars that are either reliably 
or probably identified with
EGRET sources, there is a trend (Thompson et al. 1997, 1999)
in which the
efficiency of $> 100$~MeV $\gamma$-ray production increases with
decreasing spin-down power ($\dot E \propto B^2/P^4$), or equivalently,
open field line voltage ($\Phi \propto B/P^2$).  Their efficiencies
range from 0.002 for the Crab, to of order unity for the middle-aged
pulsars Geminga and PSR~B1055--52.
As the source of \source, \psr\ would have an efficiency in accord 
with the established pattern if its distance were close to our
estimate of 3~kpc.  If, on the other hand, the distance were
as large as the DM estimate of 12 kpc, \psr\ would have to be
more efficient than either the Crab or Vela, but not so much 
more as to rule it out as the source of \source.
While \psr\ is apparently similar to the Vela pulsar in $\gamma$-ray
luminosity, its $\gamma$-ray spectrum is characterized as a
power-law of photon index $\Gamma = 2.24 \pm 0.14$
(Hartman et al. 1999).  This is steep compared to Vela and all
other EGRET pulsars except the Crab, for which $\Gamma = 2.19 \pm 0.02$.
The X-ray pulse profile of \psr\ is also reminiscent of that
of the Crab, with its two unequal peaks.

Iyudin et al. (1997) reported
the detection of a source in the 0.75--3~MeV band with COMPTEL,
coincident with \source\ but with a much larger error box.
This detection is consistent in flux with an extrapolation of the
EGRET spectrum, but it would exceed an extrapolation of the 2--10 keV
spectrum of \xray\ to 1~MeV.  The soft gamma-ray emission in this
region therefore deserves more detailed study.  \psr\ may be one
of the brightest pulsars at 1~MeV.

As a supernova remnant, G106.6+2.9 still has the peculiar property in
the radio of shell morphology, but with an extremely flat radio
spectrum ($\alpha_r \approx 0.0$, Paper~I).  Since the X-ray emission
appears to be largely confined within the radio shell, and since the
shell is too small to be the blast wave of a $10^4$ year old supernova
remnant, we conclude as was proposed in Paper~I that the radio
emission comes from a shock driven into the surrounding medium either
by the motion of the pulsar or by the expansion of the PWN.  In the
bow-shock interpretation, we can relate the now known spin-down power,
$\dot E = 2.2 \times 10^{37}$ ergs~s$^{-1}$ assumed to be carried
almost entirely by the PWN, to the velocity of the pulsar $v_p$,
the ambient density $n_{\rm H}$, and the radius of the shock $r_0$ via
$\dot E\ =\ 4\pi\,r_0^2\,c\,\rho_0\,v_p^2$.  Thus
$$\dot E =\ 
2.2\times 10^{37}\ \left ({n_{\rm H} \over 0.01}\right )
\left ({d \over 3\,{\rm kpc}}\right )^2 
\left ({v_p \over 90\,{\rm km\,s^{-1}}}\right )^2\ \ {\rm ergs\ 
s^{-1}}.$$
We measured $1.\!^{\prime}7$ for $r_0$, and we assume a relatively
low-density medium, as might be appropriate at a $z$-height of 150~pc
or in a cavity previously evacuated by a supernova explosion.
However, the production of the particle
energy spectrum needed to explain the flat radio spectrum is still not
understood theoretically.

\section {Conclusions and Further Work}

We have discovered the likely source of the $\gamma$-rays from \source,
a young and energetic 51.6~ms pulsar that has properties in accord
with those of known $\gamma$-ray pulsars.  It may be possible to prove
the association by finding the corresponding pulsed signal in the EGRET
$\gamma$-ray photons.  It is unlikely, however, that the ephemeris of
such a young pulsar is stable enough to extrapolate back with phase
coherence to the several epochs of EGRET exposure on this source.
While we have ongoing radio and X-ray observations scheduled to assist in
this analysis, considerable searching of parameter space will undoubtedly be
necessary.  Future observations
with {\it GLAST} can definitively test the identification, and can
provide an excellent pulse profile for comparison with the X-ray and
radio.  Even so, it is important to realize that since, as described in
Paper~I, we have already made a sensitive search of the entire error
circle of \source\ without finding any plausible alternative X-ray
counterpart to much fainter limits, it is more conservative to accept
the identification with \psr\ than to doubt it.  The only other object
that we are aware has been considered for identification with \source\
is a Be star/X-ray binary with an unknown rotation period
(SAX~J2239.3+6116 = 4U~2238+60, in 't Zand et al. 2000), but since that
X-ray source is $1.\!^{\circ}5$ from the $\gamma$-ray centroid, or 3
times the 95\% error radius, it is an unlikely candidate.
Meanwhile, \psr/G106.6+2.9 is sure to become a well-observed example of
a young PWN, more of which are needed to study the poorly
understood processes of particle acceleration in pulsars, spin-down
evolution, MHD flows in PWNe, and interaction with the
surrounding medium.

\acknowledgments{
We thank Christine Jordan and Benjamin Collins 
for help with reduction of radio data and X-ray data, respectively.
An anonymous referee supplied many helpful comments.
This work was supported by NASA grants NAG 5-9095 and CXO GO1--2049X
to the Columbia Astrophysics Laboratory.
}

\begin{deluxetable}{lr}
\tablenum{1}
\tablecolumns{2}
\tablewidth{0pc}
\tablecaption{Parameters of \psr\ }
\tablehead
{
\omit\hfil Parameter \hfil & \omit\hfil Value \hfil \\
}
\startdata
R.A. (J2000)        & $22^{\rm h}29^{\rm m}05.\!^{\rm s}28(7)$         \\
Decl. (J2000)       & $+61^{\circ}14^{\prime}09.\!^{\prime\prime}3(5)$ \\
Galactic Longitude                   & $106.\!^{\circ}65$          \\
Galactic Latitude                    & $2.\!^{\circ}95$            \\
Period (s)                           & 0.05162357393(6)             \\
Period Derivative (s s$^{-1}$)       & $7.827(2) \times 10^{-14}$  \\
Epoch (MJD)                          & 51980.0                     \\
Dispersion Measure (cm$^{-3}$ pc)    & 200(10)                     \\
Distance$^{\rm a}$ (kpc)             & $\sim 3$                    \\
Spin-down Luminosity (ergs s$^{-1}$) & $2.2 \times 10^{37}$        \\
Characteristic age (yr)              & 10,460                      \\
Magnetic Field (G)                   & $2.0 \times 10^{12}$        \\
\enddata
\tablenotetext{a}{See text.}
\end{deluxetable}

\end{document}